\documentclass[amsmath,amssymb,showpacs]{revtex4}
\usepackage{amsmath,amsfonts,amssymb,latexsym,extarrows}
\newtheorem{thm}{Theorem}

\renewcommand{\theequation}{\arabic{equation}}

\newcommand{\dis}{\displaystyle}

\newcommand{\bequ}{\begin{equation}}
\newcommand{\eequ}{\end{equation}}
\newcommand{\barr}{\begin{array}}
\newcommand{\earr}{\end{array}}
\newcommand{\bea}{\begin {eqnarray}}
\newcommand{\eea}{\end {eqnarray}}
\newcommand{\lb}{\label}

\begin{document}
\def \tr {\mathrm{Tr\,}}
\let\la=\lambda
\def \Z {\mathbb{Z}}
\def \Zt {\mathbb{Z}_o^4}
\def \R {\mathbb{R}}
\def \C {\mathbb{C}}
\def \La {\Lambda}
\def \ka {\kappa}
\def \vphi {\varphi}
\def \Zd {\Z ^d}
\title{Asymptotics for Some Logistic Maps and the Renormalization Group}
\author{Paulo A. Faria da Veiga}\email{veiga@icmc.usp.br.  Orcid Registration Number: 0000-0003-0739-069X.}
\author{Michael O'Carroll}\email{michaelocarroll@gmail.com}
\affiliation{Departamento de Matem\'atica Aplicada e Estat\'{\i}stica - ICMC, USP-S\~ao Carlos,\\C.P. 668, 13560-970 S\~ao Carlos SP, Brazil}
\pacs{11.15.Ha, 02.30.Tb, 11.10.St, 24.85.+p \vspace{3mm}}
\date{June 22, 2024.} 
\begin{abstract}
We explain the relation between the $r\,=\,1$ case of the logistic map $x_{i+1}\,=\, r\,x_i\,(1\,-\,x_i)$, $x_i\in\mathbb R$, $i=0,1,2,\ldots$, $r>0$ and $x_0\,\geq\,0$, and the renormalization group flow arising in the multiscale analysis of  interesting zero fixed point, asymptotic free quantum field theory models such as the ultraviolet $(1+1)$-dimensional Gross-Neveu model and QCD, and the infrared $\phi^4_4$ model . We obtain the asymptotics of the mapping, which shows an inverse power decay approach to the fixed point $x^*\,=\,0$, Gaussian fixed point, with additional logarithmic-like corrections. This asymptotic behavior is independent of the initial condition $x_0\in(0,1)$ (hence, there is no constraint for $x_0$ to be small, as usual in quantum field models), and only depends on the lowest orders in a polynomial perturbation. In asymptotic free quantum field theory, this amounts to say that knowing the renormalization group $\beta$-function expansion in the coupling constant, up to higher orders, does not improve our knowledge of the asymptotics of the coupling flow. A comparison with a similar differential equation with continuous time is made by analyzing stability of this kind of solution and higher order monomial perturbations. We also obtain the detailed asymptotics for $0\,<\,r\,<\,1$. As well, our methods can be applied when $r\,\in\,(1,3]$. It is known, but without detailed asymptotics, that all trajectories with initial condition $x_0\in(-1,1)$ converge to the fixed point $x^*\,=\,(r-1)/r$. 
For $r\,=\,2$, the super attractive case, we obtain an explicit exact solution which exhibits an exploding, non-constant exponential decay rate approach to the $x^*\,=\,(1/2)$ fixed point. Our methods include the use of iterations, a discrete version of the Fundamental Theorem of Calculus, a discrete version of the integrating factor method for first order linear ODEs and, sometimes, a scaling transformation. To obtain these results, we do {\sl not} use the traditional Banach contraction mapping theorem, which only provides an upper bound on the asymptotics. We expect that our methods can be employed to determine the asymptotics of the logistic map for a wider range of parameters, where other fixed points are present.
\end{abstract}
\maketitle
\begin{center}{\em This modest note is dedicated to the memory of our friend Krzysztof Gawedzki who\\made important contributions to mathematics and mathematical physics.}\end{center}\vspace{6mm}

\section{Introduction} \lb{intro}
For real $x_i$ and $i\in\{0,1,2,\ldots\}$, the so called {\em logistic map} is defined by \bequ\lb{logi2}
x_{i+1}\,=\, r\,x_i\,(1\,-\,x_i)\quad,\quad r>0\,.
\eequ
For $r\in(0,4]$, the mapping $x\,\mapsto\,f(x)\,=\,r\,x\,(1\,-\,x)$ takes $x\in[0,1]$ to $[0,1]$. The function $f(x)$ has the fixed points $x^*\,=\,0$ and $x^*\,=\,(r-1)/r$.

In Eq. (\ref{logi2}), by making the transformation $x_i\,=\,\beta\lambda_i$ and the identification $r\equiv \alpha$, we can define the map (in an extended domain) in the more traditional Verhulst form \cite{bookODE}, namely,
\bequ\lb{logi}
\lambda_{i+1}\,=\, \alpha\,\lambda_i\,(1\,-\,\beta\,\lambda_i)\quad,\quad  \alpha,\beta>0\,,
\eequ

This map is of a great relevance in the analysis of dynamical systems (see e.g. Refs. \cite{Wiki,Tiago} and the papers cited therein), in problems of a general mathematical context, in physics, and even in some domains of human and social sciences \cite{Chaos,Chaos2,Soc,Olver}.

The origin of this map amounts to the formulation of the famous 1845 population model by Verhulst \cite{bookODE} which incorporates a brake in the exponential growth arising in the Malthus model. If $N(t)$ denotes the population of a given specie at time $t>0$, the Verhulst nonlinear, Ricatti-type first order ordinary differential equation reads ($a,b>0$)
\bequ\lb{Verh}
\dfrac{d}{dt}{N}(t)\,= a\, N(t)\,-\,b\,[N(t)]^2\quad,\quad t>0\,,
\eequ
which is usually considered with the initial condition $N(t_0> 0)\,=\,N_0$. For $N_\infty\,=\,(a/b)$, its solution gives rise to the logistic curve (for $N_0\,<\,N_\infty$)
\bequ\lb{VerhSol}
N(t)\,=\, \dfrac{N_0\,N_\infty}{N_0\,+\,(N_\infty\,-\,N_0)\,\exp[-a(t-t_0)]}\,.
\eequ
which satisfies $N(t)=N_\infty$ as $t\nearrow\infty$, and changes the concavity from positive to negative at $N(t)\,=\,N_\infty/2$. Also, for fixed $t$, we have that (recalling $N_\infty\,=\,a/b$) 
\bequ\lb{plaw}{\cal N}(t)\,\equiv\,\lim_{a\searrow 0}\,N(t)\,=\,\dfrac{N_0}{1\,+\,N_0\,b\,(t-t_0)}\,,\eequ
such as we have a pure power law decay in $(t\,-\,t_0)$. The limit of $\left[(t\,-\,t_0)\,{\cal N}(t)\right]$ as $t\nearrow \infty$ is independent of the initial state (condition)  $N_0$ and is given by $(1/b)$, which is singular at $b\,=\,0$.

In Quantum Field Theory (QFT) \cite{Wei,Banks,Gat,GJ,Riv}, a the special case $r=1$ of the logistic map makes its appearance when the renormalization group (RG) formalism is used to perform a rigorous multiscale analysis of a QFT model. When doing so, by the RG map, the model Hamiltonian is mapped on an effective Hamiltonian model with a lower momentum cutoff (in Fourier space, we go from short distance scales to longer distances). Equivalently, the model parameters (such as the particle masses, couplings and field strengths) undergo a flow in a convenient space for the mapping. In particular, the coupling parameter flow does verify the logistic map in case of asymptotically free models, e.g. the $(1+1)$-dimensional Gross-Neveu, pure fermionic model with a quartic interaction (for the ultraviolet limit case), the  $(3+1)$-dimensional bosonic quartic massless scalar model ($\phi^4_4$) (for the infrared limit case) \cite{GK1,FMRS1,GK2,GK3,FMRS2} and the short distance behavior of $(3+1)$-dimensional Yang-Mills theories and QCD \cite{Wei,Gat,Bal,Bal2,MRS}.

In constructive QFT, even in cases of models with a small coupling, perturbation expansion is {\sl not} performed, since the perturbation series may not converge. Instead, we use other expansions (polymer, cluster, etc) to extract only the leading perturbation order for some quantity and apply a cumulant type bound to the remainder. In the rigorous multiscale method based on the RG formalism, the flow for the model parameters is analytically controlled using induction hypotheses on the parameter behaviors on the number of iteration variable (say, the `{\sl time}' $n$). To prove the induction stabilizes, it is very useful to obtain the asymptotics of the RG flow as explicit and precise as possible, at least in its leading part. In a more general context, the RG appears in the pioneering works of Refs. \cite{Kad1,Kad2,Feig1,Feig2,Feig3,Feig4}. A nice review is given in Ref. \cite{AS}.

To what concerns quantum physics, we mainly deal with the special case $\alpha\equiv r=1$ of Eq. (\ref{logi}). Concerning the logistic map, our methods are found to apply to the case $0\,<\,r\,<\,1$ and we obtain the detailed asymptotic behavior $\lim_{n\nearrow\infty}\, (\ln\lambda_n/n)\,=\,\ln r$, in Appendix A. Our methods can also be applied to the cases $1\,<\,r\,\leq\,3$ with the fixed point $x^*\,=\,(r-1)/r$, except for $r\,=\,2$, where $x^*\,=\,(1/2)$ is a fixed point. In the $r=2$ case, the fixed point is a super attractor, i.e. the derivative of $f(x)$ at the fixed point is zero. We obtain an explicit exact solution for $x_n$ which has an explosive non-constant exponential decay rate of $[2^n\,|\ln(2x_0\,-\,1)|]/n$ which grows super fast as $n\nearrow\infty$ and depends on the initial state $x_0\in(0,1)$.

It is worthwhile to stress that here, by asymptotics, we mean some control over the rate of convergence of the trajectory to a fixed point $x^*$. For example, we want to know if the decay is power-like, exponential-like, etc. We warn the reader that some authors use the term {\sl exponential law} to characterize a bound on the way the fixed point is approached and not the {\sl functional behavior} of the approach to the fixed point. A case in point is the well known Banach contraction mapping principle \cite{Chaos2,FP}. One of its consequences is that the trajectory to a fixed point is bounded by a exponential; no more details are given.

From the dynamical system point of view, the $r=1$ case corresponds to the case of a parabolic fixed point (see Refs. \cite{Chaos2,Tab} for a classification of fixed points). For $r=1$, if $f(x)=x(1-\beta x)$, then $f(0)=0$ and the derivative of $f$ at zero is $1$. We emphasize that, to the best of our knowledge, there are no results on the {\sl detailed  asymptotics} of the solution to Eq. (\ref{logi}), for $r\in (0,3]$.

In QFT, the $\alpha\,\equiv\,r\,=\,1$ case corresponds essentially to what we obtain for the flow of the coupling constant $\lambda_i$ in the case of a marginal canonical scaling. Here, one of our goals is to obtain the asymptotic behavior for this case, in terms of $n$. We obtain the asymptotics $\lim_{n\nearrow\infty}\, n\,\lambda_n\,=\, 1/\beta$ which is independent of the size of the initial state $\lambda_0\in(0,1/\beta)$ [see Theorems 1 and 2 below]. In other treatments (e.g. \cite{GK1,FMRS1,FMRS2}), $\lambda_0$ is constrained to be small. Obtaining the precise asymptotics for the leading flow of the $\lambda_i$, without this smallness restriction is also one of the achievements of this paper. 

Without solving in detail, our method to obtain asymptotics applies to the case $r\in (1,3]$. In addition to the fixed point $x^*=0$, there is another fixed point at $x^*\,=\,[(r-1)/r]$. For $r=2$, in the last section, we obtain an explicit exact formula for $x_n$ (see Eq. (\ref{solr2})). This formula exhibits a super attractive approach to the fixed point $x^*=(1/2)$ with a dependence on the initial condition.
For $r\in(1,3]$, without a detailed asymptotics, it is proved in Proposition 2.5.2 of \cite{Chaos2} that all trajectories with $x_0\in(-1,1)$ converge to the fixed point $x^*\,=\,(r-1)/r$.

Another clear conclusion emerging from our results, for $r=1$, is a stability condition that says the leading asymptotic behavior for the coupling $\lambda_i$ in QFT of asymptotic free models is determined only by the first few order coefficients in the Taylor expansion of the $\beta$ renormalization function in the renormalized coupling parameter. For example, in the perturbative approach to the spacetime two-dimensional Gross-Neveu model, it suffices to know the $\beta$ function at the one and two-loop level, i.e. the coefficients $\beta_2$ and $\beta_3$ \cite{Wet}, such that
\bequ\lb{betas}
\beta\,\equiv\,\beta(\lambda)\,=\, \beta_2\,\lambda^2\,+\,\beta_3\,\lambda^3\,+\,\beta_4\,\lambda^4\,+\,\ldots
\eequ
Knowing the coefficient $\beta_2$ and checking it has the good sign ($\beta_2\,<\,0$ is enough to treat the ultraviolet limit of the two-dimensional  Gross-Neveu model where the running coupling decreases as the momentum increases, and $\beta_2\,>\,0$ to treat the infrared limit of the massless quartic scalar model in four dimensions, where the running coupling decreases as the momentum decreases)) is enough to control the RG map (stabilizing the induction hypothesis {\sl non-perturbatively}) and proving the existence of a zero (Gaussian) fixed point, which is tautological to the existence of its continuum limit.

Using recursively the approximate asymptotic solution in terms of $\beta_2$, the knowledge of the coefficients $\beta_{j\geq 3}$, independently of their size and signs, may only introduce an extra requirement on the size of the initial condition $\lambda_0$. This condition occurs in the generalization of the logistic-type equation, which we treat in a continuum version in section \ref{sec3}, adding a cubic term. We show that the asymptotics of the solution, in this case, is determined by the quadratic term $\beta_2\lambda^2$, by a suitable restriction on the initial conditions. In doing this, the methods described in \cite{PO} are useful.


We formulate the problem and derive our results for the $r\equiv \alpha=1$ case in the next section. and make general comments in section \ref{sec4}. In the context of QFT, using the block renormalization group in \cite{Dim,Dim2,GK2,GK3}, the models $\phi^4_3$ and $QED_3$ are proved to obey ultraviolet stability bounds. In this approach, a sequence of actions and a sequence of their associated parameters are generated. The flow in the parameter space is successfully controlled applying the contraction mapping principle. For completeness, we obtain the asymptotics for the logistic map in the contractive region $0\,<\,r\equiv\alpha\,<1$  in Appendix A. Section III is devoted to determining the asymptotics for a logistic type mapping in the continuum version and we examine the stability of solutions under higher power monomial perturbations. The results we obtain are given in Theorem 3. In section IV, we make some final comments and conclusions. Since our methods can be used to determine the asymptotics for the whole range $0\,<\,\alpha\,\leq\,3$, in this last section we present the results for the $r=2$ case which develops a super attractor fixed point at $x^*=(1/2)$. 

Lastly, we emphasize that our methods apply, as well, to treat the many parameter flow problem in QFT, where there are many physical flow parameters to be controlled, such as couplings, masses, field strength, etc. This is the scenario in \cite{GK1,FMRS1,GK2,GK3,FMRS2} with a marginal coupling but also in Ref. \cite{Dim,Dim2} where the quartic coupling is relevant in three dimensions.
\section{Asymptotic Behavior for the $r=1$ Logistic Map}\lb{sec2}
Setting $\alpha\equiv r=1$ in Eq. (\ref{logi}), and recalling that $n\,=\,0,1,2,\ldots$ and $\beta>0$, we obtain
\bequ\lb{logia}
\lambda_{n+1}\,-\,\lambda_n\,=\,-\beta\,\lambda_n^2\qquad,\qquad\lambda_0\in(0,1/\beta)\,.
\eequ
We note that $\lambda_0\,=\,0$ is a fixed point and, if $\lambda_0\,=\,(1/\beta)$, then $\lambda_{k\geq 1}$ is at the fixed point zero.

From Eq. (\ref{logia}), we see that the sequence $\{\beta\lambda_n\}$ is monotone decreasing so that
\bequ\lb{ineq}
1\,>\,\beta\lambda_0\,>\,\beta\lambda_1\ldots>\beta\lambda_k\,>\,\beta\lambda_{k+1}\ldots
\eequ

We use a discrete version of the Fundamental Theorem of Calculus (FTC). To see how this works, first consider an approximation to Eq. (\ref{logia}) where $n$ is replaced by a continuum time variable $t\geq0$ and $\lambda_n$ is replaced by $\lambda(t)$. The logistic recursion is replaced by the ODE with initial value
\bequ\lb{ODE}
\dfrac{d}{dt}{\lambda}(t)\,= -\beta [\lambda(t)]^2\qquad,\qquad t>0\,\;,\,\; \lambda(0)\,=\,\lambda_0>\,0\,,
\eequ
or, simply,\bequ\lb{MalthusMinus}\dfrac{d}{dt}\left[\dfrac 1{\lambda(t)}  \right]\,=\,\beta\qquad,\qquad t\,>\,0\;,\,\;\lambda(0)\,=\,\lambda_0>\,0\,.\eequ

From the FTC, we then have
\bequ\lb{sol1}
\dfrac1{\lambda(t)}\,=\,\dfrac1{\lambda_0}\,+\,\beta\,t\,,
\eequ
or
\bequ\lb{lamt}
\lambda(t)\,=\,\dfrac{\lambda_0}{1\,+\,\beta\lambda_0t}\,\rightarrow\,\dfrac1{\beta t},\quad{\mathrm for}\,\,\,t\gg 1\,,
\eequ
which is again a pure power law decay, as in Eq. (\ref{plaw}). Of course, this asymptotics is singular in $\beta$ at $\beta=0$.

Returning to the discrete case and writing the recursion as
\bequ\lb{fraclambda}
\dfrac{\lambda_n}{\lambda_{n+1}}\,=\,\dfrac1{1\,-\,\beta\lambda_n}\,,
\eequ
the discrete version of the FTC is the telescopic sum with range on the index in $[0,n+1]$. Namely,
\bequ\barr{lll}\lb{long}
\dfrac1{\lambda_{n+1}}&=&\dfrac1{\lambda_0}\,+\, \dis\sum_{k=0}^n\, \left(  \dfrac1{\lambda_{k+1}}\,-\,\dfrac1{\lambda_{k}} \right)
\,=\,\dfrac1{\lambda_0}\,+\,\dis \sum_{k=0}^n\,\dfrac{\beta}{1\,-\,\beta\lambda_k}\vspace{2mm}\\
&=&\dfrac1{\lambda_0}\,+\,\beta\,(n+1)\,+\,\beta^2\,\dis\sum_{k=0}^n\,\lambda_k\,+\,\beta^3\,\sum_{k=0}^n\,\dfrac{\lambda_k^2}{1\,-\,\beta\lambda_k}\vspace{2mm}\\
&\equiv& \dfrac1{\lambda_0}\,+\,\beta\,(n+1)\,+\,\beta^2\,S^n_1\,+\,\beta^3\,S^n_{\geq2}\,,
\earr
\eequ
Note that, here, we also have the additional terms in $\beta^2$ and $\beta^3$, in contrast with the solution for the continuous case of Eq. (\ref{sol1}). Note also that, because of Eq. (\ref{ineq}), we can use a geometric series expansion to treat the $\beta^3$ term.

From Eqs. (\ref{long}) and (\ref{ineq}), and making $(n+1)\,\rightarrow\,n$, we immediately derive upper/lower bounds for $(1/\lambda_n)$ given by
\bequ\lb{bounds}
\dfrac1{\lambda_0}\,+\,\beta\,n \,\leq\,\dfrac1{\lambda_n} \,\leq\,\dfrac1{\lambda_0}\,+\,\dfrac{\beta\,n}{1\,-\,\beta\lambda_0}\,. 
\eequ
For the lower bound, we used $(1\,-\,\beta\lambda_k)\,<\,1$; and upper bound $(1\,-\,\beta\lambda_k)\,>\,(1\,-\,\beta\lambda_0)$, for all $k\geq1$.

In terms of $\lambda_n$, we summarize our bounds in the theorem below.
\begin{thm}\lb{thm1}
Let
\bequ\lb{laul}\barr{llc}
\lambda_n^\ell\,\equiv\,\dfrac{\lambda_0}{1\,+\, \dfrac{\beta\lambda_0}{1\,-\,\beta\lambda_0}\,n}&\longrightarrow& \dfrac{1\,-\,\beta\lambda_0}{\beta n}\quad,\quad n\gg1 \vspace{2mm}\\
\lambda_n^u\,\equiv\,\dfrac{\lambda_0}{1\,+\,\beta\lambda_0\,n}&\longrightarrow & \dfrac{1}{\beta n}\quad,\quad n\gg1\,.
\earr\eequ
Then, we have
\bequ\lb{bdthm1}
\lambda^\ell_n\,\leq\,\lambda_n\,\leq\,\lambda^u_n\,.
\eequ	
\end{thm}

The above lower and upper bounds, and their asymptotics, are not equal. But it turns out that the upper bound gives the correct asymptotics. To see this point, it suffices to bound $S_1^{n-1}$ and $S_2^{n-1}$ in Eq. (\ref{long}) noting that 
\bequ\lb{Sn}S^{n-1}_{\geq 2}\,<\,\dfrac{1}{1\,-\,\beta\lambda_0}\,S_2^{n-1}\,,\eequ
where 
$S_2^{n-1}\,\equiv\,\dis\sum_{k=0}^{n-1}\,\lambda_k^2$. The inequality follows since $(1\,-\,\beta\lambda_k)^{-1}\,\leq\, (1\,-\,\beta\lambda_0)^{-1}$, $k\geq1$. 

Using this bound, and bounding the sums over $k$ by integrals over continuously varying $k$, we see that $S_1^{n-1}$ and $S_{\geq 2}^{n-1}$ of Eq. (\ref{long}) enjoy the bounds
\bequ\lb{sbd}\barr{lll}
S_1^{n-1}&\leq&\dis\sum_{k=0}^{n-1}\,\dfrac{\lambda_0}{1\,+\,\beta\lambda_0\,k}
\,\leq\,
\lambda_0\,\left[1\,+\,\dis\int_0^{n-1}\,\dfrac1{1\,+\,\beta\lambda_0x}\,dx\right]
\,\leq\,
\lambda_0\left[1\,+\,\dfrac1{\beta\lambda_0}\,\ln\left(1\,+\,\beta\lambda_0\,n\right)\right]\,,\vspace{2mm}\\
S_{\geq2}^{n-1}&\leq&
\dfrac{1}{1\,-\,\beta\lambda_0}\;\dis\sum_{k=0}^{n-1}\,\lambda^2_k
\,\leq\,
\dfrac{\lambda_0^2}{1\,-\,\beta\lambda_0}\,
\dis\sum_{k=0}^{n-1}\,\dfrac1{\left(1\,+\,\beta\lambda_0\,k\right)^2}
\,\leq\,
\dfrac{\lambda_0^2}{1\,-\,\beta\lambda_0}\,\;\left[1\,+\,\,\dis\int_0^{n-1}\dfrac1{\left(1\,+\,\beta\lambda_0\,x\right)^2}\,dx\right]\vspace{2mm}\\
&\leq & 
\dfrac{\lambda_0^2}{1\,-\,\beta\lambda_0}\,\left[1\,+\,\dfrac{n-1}{1\,+\, \beta\lambda_0\, (n-1)}\,\right]\,\leq\,\dfrac{\lambda_0}{1\,-\,\beta\lambda_0}\,\left(\lambda_0\,+\,\dfrac1\beta\right)\,,
\earr
\eequ
from which we see that the bound for $S_1^{n-1}$ blows up logarithmically and $S_2^{n-1}$ remains bounded for large $n$.

Again, making $(n+1)\rightarrow n$ in Eq. (\ref{long}), and dividing by $n$, we have 
\bequ\lb{asymplim}\barr{lll}
\dfrac1{n\,\lambda_n}\,-\,\beta&=& \dfrac{\beta^2}{n}\,S_1^{n-1}\,+\,\dfrac{\beta^3}n\,S^{n-1}_{\geq 2}\vspace{2mm}\\
&\leq&\dfrac{\beta^2\lambda_0}{n}\,\left[1\,+\,\dfrac1{\beta\lambda_0}\,\ln\left(1\,+\,\beta\lambda_0\,n\right)\right]\,+\,\dfrac{\beta^3\lambda_0}{n\left(1\,-\,\beta\lambda_0\right)}\;\left(\lambda_0\,+\,\dfrac1\beta\right).
\earr\eequ

Taking the $n\nearrow\infty$ limit, the r.h.s. of the first line of Eq. (\ref{asymplim}) gives zero and we obtain the correct asymptotics which is stated in the theorem below and which agrees with the Verhulst continuum case.
\begin{thm}\lb{thm2}
	The following limit holds
	\bequ\lb{nlim}\lim_{n\nearrow\infty}\,n\,\lambda_n\,=\,\dfrac1\beta\,.\eequ
\end{thm}

To finish, we can ask what we can say about the behavior of
\bequ\lb{solution}
\lambda_n\,=\,\dfrac{\lambda_0}{1\,+\,\beta\lambda_0n\,+\,\beta^2\lambda_0S_1^{n-1}\,+\,\beta^3\lambda_0S_{\geq 2}^{n-1}}\,,
\eequ
and how it compares with the solution $\lambda(t)$ of Eq. (\ref{lamt}) of the continuum model. In Eq. (\ref{sbd}), we obtained an upper bound on $S^{n-1}_1$ behaving like $\ln n$. We can use the lower bound in Theorem \ref{thm1} to show that
$S^{n-1}_1$ is also bounded below by a $\ln n$ behavior. Thus, in addition to the $(1/n)$ falloff, there is also a logarithmic-like correction. The term $\beta^3\lambda_0S^{n-1}_{\geq2}$ is bounded by an $n$-independent constant.

In the continuous ODE version, the solution is given in Eq. (\ref{lamt}) as $\lambda_0/[1\,+\,\beta\,\lambda_0\,n]$ such as the falloff is a pure power law which, in the discrete case above, we have logarithmic-like corrections due to the presence of the term $\beta^2\lambda_0\,S_1^{n-1}$.
\section{Asymptotics for a Generalization of a Logistic Type Model}\lb{sec3}

Here we determine the asymptotics for a class of autonomous first order nonlinear ODE's of the form
\bequ\lb{nlode}
\dfrac{d\lambda(t)}{dt}\,=\,F(\lambda(t))\qquad,\qquad t>0\quad,\quad \lambda(t)\in\mathbb R\,,
\eequ
where $F(\lambda)\in\mathbb R$ is a nonlinear part, and with initial condition  $\lambda(0)\,=\,\lambda_0$. Our methods can also be used to determine the asymptotics of some of the discrete versions of this type.

If the function $F(\lambda)$ is continuously differentiable, then local existence and uniqueness of the solution of Eq. (\ref{nlode}) is assured \cite{Chaos2,HS,HSD,Ince}. We assume these conditions are enforced from now on. Distinct initial conditions give rise to non-intersecting trajectories in the $\lambda\,\times\,t\,$ plane. A trajectory may run off to infinity in a finite time.

As is well-known, by separation of variables, the solution is formally given by
\bequ\lb{sol}
G(\lambda)\,-\,G(\lambda_0)=\,\dis\int_{\lambda_0}^{\lambda} \dfrac{d\lambda}{F(\lambda)}\,=\,\int_o^t\,dt\,=\,t \quad\Longrightarrow\quad G(\lambda)\,=\,G(\lambda_0)\,+\,t\,,\eequ and the explicit solution is
\bequ\lb{exsol}\lambda(t)\,=\,G^{-1}\left(G(\lambda_0)\,+\,t\right)\,.
\eequ

We will consider $F(\lambda)$ as occurs in the RG formalism, multiscale analysis of many lattice and continuum QFT models. We first take:
\bequ\lb{cubic}
F(\lambda)\,=\, -\beta\,\lambda^2\,-\, \beta_3\,\lambda^3\qquad ,\qquad \beta>0\:,\:\beta_3\in\mathbb R\,,
\eequ
and determine the $t\nearrow \infty$ asymptotics. The method will make it clear how to treat a wider class of $F$'s of Eq. (\ref{nlode}).

Before considering our specific $F(\lambda)$, we make some comments on the procedure to analyze and obtain a qualitative picture of the trajectories $\lambda(t)$. By unicity, the trajectories do not cross.
First, we determine the set of distinct zeroes of $F(\lambda)$. Denote these zeroes by $\lambda_1^*\,<\, \lambda_2^*\,\ldots$ These points are called the equilibrium points and they give rise to the constant trajectories, i.e. $\lambda_j(t)\,=\,\lambda_j^*$, with $\,t\,>\,0$.
Next, we can independently look at the open interval between two consecutive equilibrium points. For example, let us focus our attention on the interval $(\lambda_1^*\,,\,\lambda_2^*)$. In this interval, $F(\lambda)$ has the same sign. If $F(\lambda)\,>\,0$
(respectively, $F(\lambda)\,<\,0$), then the trajectory tends to the equilibrium point $\lambda_2^*$ (respectively, $\lambda_1^*$), as $t\,\nearrow\,\infty$. This qualitative picture does not give a quantitative information on the rate of convergence to the equilibrium point, i.e. the asymptotics.

We now return to the specific case of the cubic $F(\lambda)$ of Eq. (\ref{cubic}). If $\beta\,=\,0$, the explicit solution is
\bequ\lb{sol3}
\lambda(t)\,=\, \dfrac 1{\left[\dfrac1{\lambda_0^2}\,+\,2\beta_3\,t\right]^{1/2}}\,.
\eequ
For $\beta_3\,>\,0$, the asymptotics is $\lambda(t)\,\rightarrow\,\left[2\,\beta_3\,t\right]^{-1/2}$, which is independent of $\lambda_0$. We also note that the decay is slower than the linear decay $[1/(\beta\,t)]$ of the case $\beta\,>\,0$ and $\beta_3\,=\,0$. If $\beta_3\,<\,0$, $\lambda(t)$ becomes infinite in a finite time $t\,=\,\left[ 2\,|\beta_3|\,\lambda_0^2 \right]^{-1}$.

Now, what is the effect of the sign of $\beta_3$ on the solution and its asymptotics in the case of Eq. (\ref{cubic}), for $\lambda_0\,>\,0$? 
We explain intuitively what is going on. For small $\lambda_0$, the trajectory is decaying to zero so that the $\lambda^2$ term of Eq. (\ref{cubic}) dominates the $\lambda^3$ term. Thus, the solution becomes smaller and the asymptotics is determined by the $-\beta\,\lambda^2$ term.

Considering $F(\lambda)$ of Eq. (\ref{cubic}), the asymptotics for both signs of $\beta_3$ is easily obtained by using the FTC to obtain an integral representation for $[1/\lambda(t)]$ and qualitative knowledge of the solution permits us to obtain the precise asymptotics. We consider the two cases separately, $\beta_3\,>\,0$ and
$\beta_3\,<\,0$.

For $\beta\,>\,0$ and $\beta_3\,>\,0$,  we write Eqs. (\ref{nlode}) and (\ref{cubic}) as
\bequ\lb{inv}
\dfrac d{dt}\,\left[\dfrac1{\lambda(t)}\right]\,=\,\beta\,+\,\beta_3\lambda(t)\,.
\eequ

The equilibrium points are $\lambda^*_1\,=\,0$ and $\lambda^*_2\,=\, -\,\beta/\beta_3$.

We only consider $\lambda_0\,>\,0$. Applying the FTC gives
\bequ\lb{1st}
\dfrac1{\lambda(t)}\,=\,\dfrac1{\lambda_0}\,+\,\beta\,t\,+\,\beta_3\,\dis\int_0^t\,\lambda(u)\,du\,\geq\,\dfrac1{\lambda_0}\,+\,\beta\,t\,,
\eequ
that is
\bequ\lb{ubd}
\lambda(t)\,\leq\,\frac1{\lambda_0^{-1}\,+\,\beta t}\,,
\eequ
where the inequality follows since $F(\lambda)\,<\,0$, so that the solution $\lambda(t)$ is positive and monotone decreasing. Thus, $\int_0^t\, \lambda(u)\,du\,>\,0$.

The integral term of Eq. (\ref{1st}) has the bound
\bequ\lb{3bd}
\beta_3\,\dis\int_0^t\,\lambda(u)\,du\,\leq\,\beta_3\,\dis\int_0^t\,\frac1{\lambda_0^{-1}\,+\,\beta u}\,du\,=\,\dfrac{\beta_3}\beta\,\ln\left(1\,+\,\beta\,\lambda_0\,t\right)\,.
\eequ 

Dividing Eq. (\ref{1st}) by $t$ and taking the $t\nearrow\infty$ limit gives
\bequ\lb{tinfty}
\lim_{t\nearrow\infty}\, \dfrac1{t\lambda(t)}\,=\,\beta\,,
\eequ
which agrees with our preceding similar results.
Note that the limit in Eq. (\ref{tinfty}) is also independent of $\lambda_0\,>\,0$.

For $\beta\,>\,0$ and $\beta_3\,<\,0$, the equilibrium points are $0$ and $(-\beta/\beta_3)\,>\,0$. We restrict $\lambda_0$ to the interval $\left(0,(-\beta/\beta_3)\right)$. The representation of Eq. (\ref{1st}) also hold in this case. With the above restriction on $\lambda_0$, $\lambda(t)$ is positive and monotone decreasing as $F(\lambda)\,<\,0$.

For $0\,<\,f\,<\,1$, we now write 
\bequ\lb{lambda0} \lambda_0\,=\,-\frac{\beta}{\beta_3}\,(1\,-\,f)\,.\eequ
Applying the FTC once more, we have
\bequ\lb{bd4}\barr{lll}
\dfrac1{\lambda(t)}&=&\dfrac1{\lambda_0}\,+\,\beta t\,+\,\beta_3\,\dis\int_0^t\,\lambda(u)\,du\vspace{2mm}\\
&\geq&\dfrac1{\lambda_0}\,+\,\beta\,t\,-\,|\beta_3|\,\lambda_0\,t\,,\earr
\eequ or
\bequ\lb{bd44}
\lambda(t)\,\leq\,\,\dfrac1{\dfrac1{\lambda_0}\,+\,\left(\beta\,-\,|\beta_3|\,
	\lambda_0\right)\,t}\,,
\eequ
where we use the fact that $\lambda(t)$ is positive and monotone decreasing. We have also restricted $\lambda_0$ to the interval $(0,\,-\beta/\beta_3\,>\,0)$.

The integral term of Eq. (\ref{bd4}) is bounded by
\bequ\lb{bd4b}\barr{lll}
|\beta_3\,I(t)|&\leq&
\dfrac{|\beta_3|}{\beta\,f}\,\ln\left(1\,+\,\dfrac{(1-f)\,\beta^2\,f}{|\beta_3|}\,t  \right)\,.
\earr
\eequ
As before, dividing Eq. (\ref{1st}) by t and taking $\lim_{t\nearrow\infty}$ gives the asymptotics of Eq. (\ref{tinfty}). Hence, $\beta[\lambda(t)]^2$ determines the asymptotics.

We summarize these findings in the following theorem that concludes this section.
\begin{thm}
The solution of the ODE
\bequ\lb{EDO2}
\dfrac{d\lambda(t)}{dt}\,=\,-\,\beta\,\lambda^2\,-\,\beta_3\,\lambda^3\qquad;\qquad t>0\quad,\quad \lambda(t=0)\,=\,\lambda_0\quad,\quad\beta>\,0\,,
\eequ
has the asymptotics
\bequ\lb{asymp}\lim_{t\nearrow\infty}\,t\,\lambda(t)\,=\,\dfrac1\beta\,,\eequ for $\beta_3\,>\,0$ and $\lambda_0\,>\,0$ or for $\beta_3\,<\,0$ and $\lambda_0\,\in\,\left(0,-(\beta/\beta_3)\,>\,0\right)$.
\end{thm}
\section{Final Comments}\lb{sec4}
We determined the asymptotics of the logistic map (equation) $x_{n+1}\,=\, r\,x_n\,(1\,-\,x_n)$, $n=0,1,2,\ldots$, $x_i\in\mathbb R$,  for some values of $r\,>\,0$. For $r\in(0,4]$, the function $f(x)=rx(1-x)$ takes the interval $[0,1]$ to [0,1] and has fixed points at $x^*\,=\,0,[(r-1)/r]$. Our results on the asymptotics are obtained by performing iterations and using the method of discrete fundamental theorem of calculus, discrete integrating factor method to solve first order linear ODEs and, eventually, also performing scaling transformations. In particular, we treated the cases $r\in(0,1]$ explicitly but our methods apply to the case  $r\in(1,3]$, except for $r\,=\,2$. For $r=1$, the decay to the $x^*\,=\,0$ fixed point is like $1/n$ and is independent of the initial condition $x_0$. For $0\,\leq\,r\,<\,1$, the decay to $x^*\,=\,0$ is exponential-like for large $n$, with a decay rate of $|\ln r|$, independent of $x_0$. Our methods also allow us to determine corrections to the asymptotic behavior. For $r\,=\,1$, the correction to the $1/n$ decay is logarithmic-like. For $0\,<\,r\,<\,1$, the correction to the exponential decay rate $|\ln r|$ is roughly ${\mathcal O}(1/n)$.

For $r\,=\,2$, there is an additional fixed point $x^*\,=\,[(r\,-\,1)/r]\,=\,(1/2)$, and a decay dependence on $x_0$ arises. The exponential decay rate to $x^*\,=\,(1/2)$ explodes and is non constant, giving a super fast exponential decay to $x^*\,=\,(1/2)$. This is obtained by expanding the map around the $x^*=(1/2)$ fixed point which gives rise to a logistic-type equation with $r$ replaced by an effective $r$. Letting $r_{eff}\,\equiv\,(2\,-\,r)$ denote this effective $r$, we have $|r_{eff}|\,\leq\,|2-r|\,<\,1$. Namely, by making the change of variables $x_n\,=\,(1/2)\,+\,(y_n/2)$, which includes a scaling, we arrive at the following explicit solution
\bequ\lb{solr2}
x_n\,-\,\dfrac12\, =\,-\,\dfrac12\, 2^{2^n}\,\left( x_0\,-\,\dfrac12  \right)^{2^n}\,=\,-\,\dfrac12\,\exp\left\{-n\,\left[\dfrac{2^n}n\,\left|\ln [2\,\left(x_0\,-\,1/2\right)]\right|\right]\right\}\,\quad;\quad n\geq1\quad,\quad x_0\in\,(0,1)\,.
\eequ
Alternatively, instead of the scaling, we can use a discrete version of the integrating factor method to solve first order linear ODEs, and iteration, to arrive at Eq. (\ref{solr2}). Here, we see a non-constant exponential decay rate which depends on the initial condition $x_0$ and grows super fast as $2^n/n$. This is a super attractive case!  When $n\nearrow \infty$, $x_n\,\nearrow\,(1/2)$. This case is qualitatively described in dynamical systems books like Ref. \cite{Chaos2}.

For $r\,>\,3$, the linear part of the equation becomes expansive. From the mathematical viewpoint, it would be interesting to explore our method coupled possibly with other methods in this region of values for $r$. The tools we employed here should be helpful to analyze the logistic map with $r$ in the interval $3\,<\,r\,\leq\,4$.

Besides, from the mathematical physics point of view, we recognized that the $r=1$ logistic map occurs in rigorous multiscale analysis of asymptotic free quantum field theory models using the renormalization group. 

The RG multiscale formalism used in e.g. Refs. \cite{GK1,FMRS1}, the analytical control of the RG mapping is usually based on induction hypotheses on the behavior of the flow of the model parameters under the RG mapping. The control of the flow, and then of the induction, is more easily done if we know the asymptotics of the RG map more precisely. This was the main motivation for the present paper. In \cite{GK1,FMRS1}, the situation is a little trickier than the single logistic map treated here. The RG flow includes the map for more than one parameter (there are e.g. the flows for the coupling $\lambda$, particle masses and  field strength parameters). The analysis of the $\lambda$ flow is dominated by the above arguments and leads to the proof of a zero (Gaussian) fixed point.

In closing, we stress that our methods apply as well to treat the whole parameter flow question in QFT with more than one physical parameter. This is the scenario in \cite{GK1,FMRS1,GK2,GK3,FMRS2} with a marginal coupling but also in Refs. \cite{Dim,Dim2} where the quartic coupling is relevant in three dimensions.
\vspace{3mm}
\appendix{\begin{center}{\bf APPENDIX A: Logistic Map Asymptotics for $0\,<\,r\,<\,1$}\end{center}}
\lb{appA}
\setcounter{equation}{0}
\setcounter{lemma}{0}
\setcounter{thm}{0}
\setcounter{rema}{0}
\renewcommand{\theequation}{A\arabic{equation}}
\renewcommand{\thethm}{A\arabic{thm}}
\renewcommand{\thelemma}{A\arabic{lemma}$\:$}
\renewcommand{\therema}{{\em{\bf A\arabic{rema}$\:$}}}\vspace{.5cm}
We reconsider the logistic map in the form given in Eq. (\ref{logi2}). With $\lambda_0\in(0,1)$, we have, with $0\,<\,r\,<\,1$,
\bequ\lb{logi22}
\lambda_{n+1}\,=\,f(\lambda_n)\,\equiv\,r\,\lambda_n\,\left(1\,-\,\lambda_n \right)\,,
\eequ
or
\bequ\lb{logi22b}
\lambda_{n+1}\,-\,\lambda_n\,=\,(r\,-\,1)\,\lambda_n\,-\,r\,\lambda^2_n\,,
\eequ
so that $\lambda_n\,>\,0$ and the sequence of the $\lambda_j$ is monotone decreasing. 

From Eq. (\ref{logi22}), 
\bequ\lb{123}
\lambda_{n+1}\,\leq\,r\,\lambda_n\,=\,r\,\lambda_{n-1}\,\left(1\,-\,\lambda_{n-1}\right)\,\leq\,r^2\,\lambda_{n-1}\,\ldots\,,
\eequ
or
\bequ\lb{124}
\lambda_{n+1}\,\leq\,r^{n+1}\,\lambda_0\,=\,e^{-\,(n+1)\,|\ln r|}\,\lambda_0\,,
\eequ
an upper bound on $\lambda_{n+1}$.

Considering $\ln \lambda_n$, we have the recursion
\bequ\lb{125}
\ln \lambda_{n+1}\,=\,\ln r\,+\,\ln \left[\lambda_n\,\left(1\,-\,\lambda_n\right)\right]\,,
\eequ
or
\bequ\lb{125}
\ln \lambda_{n+1}\,-\,\ln\lambda_n\,=\,\ln r\,+\,\ln\left(1\,-\,\lambda_n\right)\,.
\eequ

Applying the telescoping method (the discrete FTC), we get
\bequ\lb{126}\barr{lll}
\ln \lambda_{n+1}&=&\dis\sum_{k=0}^{n}\,\left[\ln \lambda_{k+1}\,-\,\ln \lambda_{k}\right]\,+\,\ln\lambda_0\vspace{2mm}\\&=&\dis\sum_{k=0}^{n}\,\left[\ln {r}\,+\,\ln\left(1\,-\, \lambda_{k}\right)\right]\,+\,\ln\lambda_0\,.
\earr
\eequ

To obtain the $n\nearrow \infty$ limit of $[\ln\lambda_n/n]$, where
\bequ\lb{127}
\dfrac 1n\,\ln \lambda_{n}\,=\,\ln r\,+\,\dfrac 1n\,\ln \lambda_{0}\,+\,\dfrac1n\,
\dis\sum_{k=0}^{n-1}\,\ln \left(1\,-\,\lambda_k\right)\,,\eequ
we need to control the last sum \bequ\lb{Sn}S_{n-1}\,\equiv\,\sum_{k=0}^{n-1}\,\ln \left(1\,-\,\lambda_k\right)\,.\eequ

To bound $S_{n-1}$ of Eq. (\ref{Sn}), we note that the function $g(\lambda)\,\equiv\,\left[-\ln \left(1\,-\,\lambda\right)\right]$ is positive and convex, for $\lambda\in[0,1)$. The graph of $g(\lambda)$ is below the straight line going from the origin to $(\lambda_0,-\ln(1\,-\,\lambda_0))$. Hence, we have the bound
\bequ\lb{bdd}
g(\lambda)\,\leq\, -\,\dfrac{\lambda}{\lambda_0}\,\ln\left(1\,-\,\lambda_0\right)\,.
\eequ
Thus, 
\bequ\lb{bdSn}\barr{lll}|S_{n-1}|&\leq&\dfrac {|\ln\left(1\,-\,\lambda_0\right)|}{\lambda_0}\,\dis\sum_{k=0}^{n-1}\,\lambda_k
\,\leq\,|\ln\left(1\,-\,\lambda_0\right)|\:\,\dfrac1{1\,-\,e^{-|\ln r|}}\,,
\earr
\eequ
where we used the bound of Eq. (\ref{124}).

In Eq. (\ref{bdSn}), the bound is independent of $n$ and we finally have the asymptotics
\bequ\lb{finalbd}
\lim_{n\nearrow\infty}\,\dfrac{\ln \lambda_n}n\,=\,\ln r\,.
\eequ

As is well-known, since
\bequ\lb{fact}
f(x)\,-\,f(y)\,=\,r\,(x-y)\,[1\,-\,(x+y)]\,,
\eequ
and $|f(x)\,-\,f(y)|\,\leq\,r\,|x-y|$, for $x,y\,\in\,[0,1]$, the contraction mapping principle \cite{Chaos2,FP} can be applied to Eq. (\ref{logi22}) with $\lambda_0$ in the closed interval $[0,1]$ and, as a consequence, the upper bound of Eq. (\ref{124}) is obtained, but {\sl not} the asymptotics of Eq. (\ref{finalbd}).
\begin{acknowledgements}
	We would like to thank the anonymous referees for suggestions, and Prof. T. Pereira and Prof. D. Smania for discussions.\vspace{3mm}
\end{acknowledgements}

\end{document}